\documentclass[aps,prb,twocolumn,showpacs,preprintnumbers,amsmath,amssymb,superscriptaddress]{revtex4}%

\usepackage{graphicx}%
\usepackage{dcolumn}
\usepackage{amsmath}
\usepackage{color}
\usepackage{multirow}

\begin{document}


\title{Superconducting and magnetic properties of Sr$_3$Ir$_4$Sn$_{13}$}

\author{P.~K.~Biswas}
\email[Corresponding author: ]{pabitra.biswas@psi.ch}
\affiliation{Laboratory for Muon Spin Spectroscopy, Paul Scherrer Institute, CH-5232 Villigen PSI, Switzerland}

\author{A.~Amato}
\affiliation{Laboratory for Muon Spin Spectroscopy, Paul Scherrer Institute, CH-5232 Villigen PSI, Switzerland}
\author{R.~Khasanov}
\affiliation{Laboratory for Muon Spin Spectroscopy, Paul Scherrer Institute, CH-5232 Villigen PSI, Switzerland}
\author{H.~Luetkens}
\affiliation{Laboratory for Muon Spin Spectroscopy, Paul Scherrer Institute, CH-5232 Villigen PSI, Switzerland}
\author{Kefeng~Wang}
\affiliation{Condensed Matter Physics and Materials Science Department, Brookhaven National Laboratory, Upton, New York 11973, USA}
\author{C.~Petrovic}
\affiliation{Condensed Matter Physics and Materials Science Department, Brookhaven National Laboratory, Upton, New York 11973, USA}
\author{R.~M.~Cook}
\affiliation{Physics Department, University of Warwick, Coventry, CV4 7AL, United Kingdom}
\author{M.~R.~Lees}
\affiliation{Physics Department, University of Warwick, Coventry, CV4 7AL, United Kingdom}
\author{E.~Morenzoni}
\email[Corresponding author: ]{elvezio.morenzoni@psi.ch}
\affiliation{Laboratory for Muon Spin Spectroscopy, Paul Scherrer Institute, CH-5232 Villigen PSI, Switzerland}
\date{\today}

\begin{abstract}
Magnetization and muon spin relaxation or rotation ($\mu$SR) measurements have been performed to study the superconducting and magnetic properties of Sr$_3$Ir$_4$Sn$_{13}$. From magnetization measurements the lower and upper critical fields of Sr$_3$Ir$_4$Sn$_{13}$ are found to be 81(1) Oe and 14.4(2) kOe, respectively. Zero-field $\mu$SR data show no sign of any magnetic ordering or weak magnetism in Sr$_3$Ir$_4$Sn$_{13}$. Transverse-field $\mu$SR measurements in the vortex state provided the temperature dependence of the magnetic penetration depth $\lambda$. The dependence of $\lambda^{-2}$ with temperature is consistent with the existence of single $s$-wave energy gap in the superconducting state of Sr$_3$Ir$_4$Sn$_{13}$ with a gap value of 0.82(2) meV at absolute zero temperature. The magnetic penetration depth at zero temperature $\lambda(0)$ is 291(3)~nm. The ratio $\Delta(0)/k_{\rm B}T_{\rm c}=2.1(1)$ indicates that Sr$_3$Ir$_4$Sn$_{13}$ should be considered as a strong-coupling superconductor.
\end{abstract}

\maketitle


Recently, ternary intermetallic stannide compounds, $R$$_3$Ir$_4$Sn$_{13}$, where $R$ = Ca, Sr, etc., have attracted renewed interest because of the coexistence of superconducting and charge density wave states and the possible presence of  pressure induced quantum structural phase transitions.~\cite{Klintberg} Sr$_3$Ir$_4$Sn$_{13}$ superconducts below the transition temperature ($T_{\rm c}$) of 5~K, whereas the sister compound, Ca$_3$Ir$_4$Sn$_{13}$, becomes superconducting below 7~K.~\cite{Remeika,Espinosa} Since the Ca atom is smaller in size than the Sr, the substitution of Ca on the Sr site corresponds to applying positive pressure, which then enhances the $T_{\rm c}$ in the sister compound. This trend seems to continue in Ca$_3$Ir$_4$Sn$_{13}$ even for physical pressure. Under hydrostatic pressure, the $T_{\rm c}$ of Ca$_3$Ir$_4$Sn$_{13}$ increases to 8.9~K in 4 GPa and then falls for higher pressures.~\cite{Klintberg} An increase of $T_{\rm c}$ with increasing pressure has also been observed for Sr$_3$Ir$_4$Sn$_{13}$, a behavior at variance with that of the majority of BCS-like superconductors. In Sr$_3$Ir$_4$Sn$_{13}$, an anomaly at $T^*\approx147$~K has been detected in resistivity and susceptibility measurements. A similar anomaly at $T^*\approx33$~K has also been found in the compound Ca$_3$Ir$_4$Sn$_{13}$ and initially attributed to ferromagnetic (FM) spin fluctuations, related to the superconductivity appearing at lower temperature.~\cite{Yang} Later, single crystal x-ray diffraction studies~\cite{Klintberg} showed that the $T^*$ anomaly in Sr$_3$Ir$_4$Sn$_{13}$ is produced by a second-order superlattice transition from a simple cubic parent phase, the $I$-phase, to a superlattice structure, the $I'$-phase, with a lattice parameter twice that of the $I$-phase. It has been further argued that this superlattice transition is associated with a charge density wave (CDW) transition of the conduction electron system. Hall and Seebeck coefficient indicate gap opening and significant Fermi surface reconstruction at $T^*$ in Ca$_3$Ir$_4$Sn$_{13}$.~\cite{Wang} Whereas a low $\frac{T_{\rm c}}{T_F}$ ratio from thermoelectric data shows a weakly correlated superconductor,~\cite{Wang} other parameters such as the Wilson and Kadowaki-Woods ratios close to those of heavy fermions have been taken as indicative of stronger correlated system.~\cite{Yang}
Specific heat measurements on $R$$_3$$T$$_4$Sn$_{13}$ ($R$ = Sr,La, $T$=Ir, Rh) and Ca$_3$Ir$_4$Sn$_{13}$ ~\cite{Kase,Yang} suggest nodeless superconductivity and strong coupling, but thermal conductivity data on Ca$_3$Ir$_4$Sn$_{13}$  did not exclude some gap anisotropy, or multiple isotropic gaps with different magnitudes.~\cite{Zhou}
Recent $\mu$SR study on Ca$_3$Ir$_4$Sn$_{13}$ \cite{Gerber} determined a very high gap-to-$T_{\rm c}$ ratio value $\Delta(0) / (k_{\mathrm{B}} T_c)$ = 5, which is unusually large even for a very strongly coupled BCS superconductor and much larger than the 2.53 inferred from macroscopic measurements.~\cite{Hayamizu} To obtain a better insight into the nature of the superconducting and magnetic state of these intermetallic stannide compounds and resolve the discrepancy between the experimental results, we have performed $\mu$SR and  magnetization measurements on Sr$_3$Ir$_4$Sn$_{13}$, which is isoelectronic to Ca$_3$Ir$_4$Sn$_{13}$. 

From the magnetization measurements we extract the lower and upper critical fields of Sr$_3$Ir$_4$Sn$_{13}$ of 81(1) Oe and 14.4(2) kOe, respectively. Zero-field (ZF)-$\mu$SR results find no evidence of any magnetism in Sr$_3$Ir$_4$Sn$_{13}$. Transverse-field (TF)-$\mu$SR results show that, at low temperature, the superfluid density $\rho_s\propto\lambda^{-2}$ ($\lambda$ the magnetic penetration depth) becomes temperature independent, which is consistent with a fully gapped superconducting state. The $\rho_s(T)$ can be well fitted with a single $s$-wave gap model with a gap value of 0.82(2) meV at absolute zero temperature. This gives a gap to $T_{\rm c}$ ratio of 2.1(1). The absolute value of the magnetic penetration depth is determined to be $\lambda(0)=291(3)$~nm.


Single crystal samples of Sr$_3$Ir$_4$Sn$_{13}$ were prepared and characterized as described in Ref.~\onlinecite{Wang}. Magnetization measurements were performed using an Oxford Instruments Vibrating Sample Magnetometer (VSM). The TF- and ZF-$\mu$SR experiments were carried out at the Dolly instrument($\pi$E1 beam line) and at the new High field and Low Temperature  instrument (HAL-9500) at the $\pi$M3 beam line of the Paul Scherrer Institute (Villigen, Switzerland). The sample was cooled to the base temperature in zero field for the ZF-$\mu$SR experiments and in 500~Oe for the TF-$\mu$SR experiments. Typically $\sim10$~million muon decay events were collected for each spectrum. The ZF- and TF-$\mu$SR data were analyzed by using the free software package MUSRFIT.~\cite{Suter}


\begin{figure}[htb]
\includegraphics[width=1.0\linewidth]{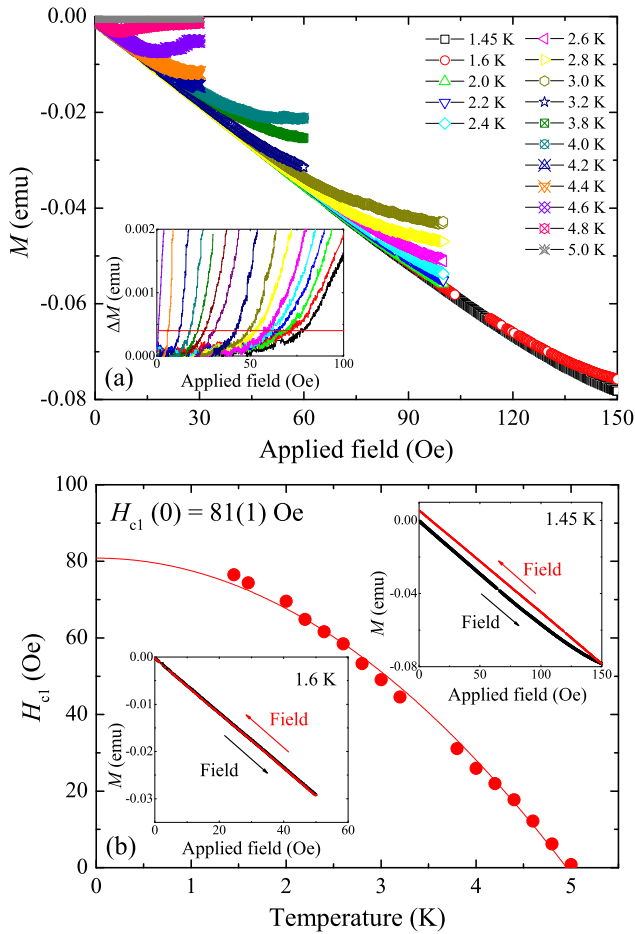}
\caption{(Color online) (a) Low-field virgin magnetization curves of Sr$_3$Ir$_4$Sn$_{13}$, collected at different temperatures. The inset shows the departure from linearity, $\Delta{M}$, calculated by subtracting the linear fit from each of the $M$ vs. $H$ curves. (b) The temperature dependence of $H_{\rm c1}$. The solid line is a quadratic fit to the data using Eq.~\ref{eq:quadratic_fit}. The insets show the $M$ vs. $H$ curves up to various maximum applied field values (above and below the $H_{\rm c1}$ value) that validate our estimate of $H_{\rm c1}$.}
 \label{fig:SrIrSn_Hc1}

\end{figure}

Figure~\ref{fig:SrIrSn_Hc1}(a) shows the virgin magnetization $M$ versus applied field $H$ curves for Sr$_3$Ir$_4$Sn$_{13}$ collected at various temperatures. These measurements were performed to obtain the temperature dependence of lower critical field $H_{\rm c1}$ by estimating the first deviation from linearity in each of the $M$ vs. $H$ curves. To do this, a linear fit to the data was made between 0 to 10 Oe. The departure from linearity, $\Delta{M}$ was calculated by subtracting the linear fit from each of the $M$ vs. $H$ curves (see the inset of Fig.~\ref{fig:SrIrSn_Hc1}(a)). The temperature dependence of $H_{\rm c1}$ was then obtained by using the criteria $\Delta{M}=4\times{10}^{-4}$ emu, indicated in the figure as a solid horizontal line. Fig.~\ref{fig:SrIrSn_Hc1}(b) shows $H_{\rm c1}(T)$ of Sr$_3$Ir$_4$Sn$_{13}$. The solid line is a quadratic fit to the data using
\begin{equation}
H_{\rm c1}(T)=H_{\rm c1}(0)\left\{1-\left(\frac{T}{T_{\rm c}}\right)^2\right\}.
 \label{eq:quadratic_fit}
\end{equation}
Although, the quadratic fit to the $H_{\rm c1}(T)$ data does not look very good but it certainly gives a fair estimate for $H_{\rm c1}(0)$. We obtain $H_{\rm c1}(0)$=81(1) Oe. We checked our $H_{\rm c1}(0)$ estimation by following procedure. In a first attempt, at 1.45~K, we increased the applied field up to 150~Oe (well above $H_{\rm c1}(1.45)=74.4$~Oe, estimated from the fitted curve) and then decreased the field back to zero. The presence of hysteresis in the data (see the top inset of Fig.~\ref{fig:SrIrSn_Hc1}(b)) implies that in this case some flux lines have entered the sample and that the sample has crossed the $H_{\rm c1}$ limit to the mixed state. In a second attempt, we increased the applied field up to 50~Oe at 1.6~K (below $H_{\rm c1}(1.6)=72.3$~Oe). In this case, the curve is completely reversible (see the bottom inset of Fig.~\ref{fig:SrIrSn_Hc1}(b)) confirming that the applied field is well below the $H_{\rm c1}$ value at 1.6~K.

\begin{figure}[htb]
\includegraphics[width=1.0\linewidth]{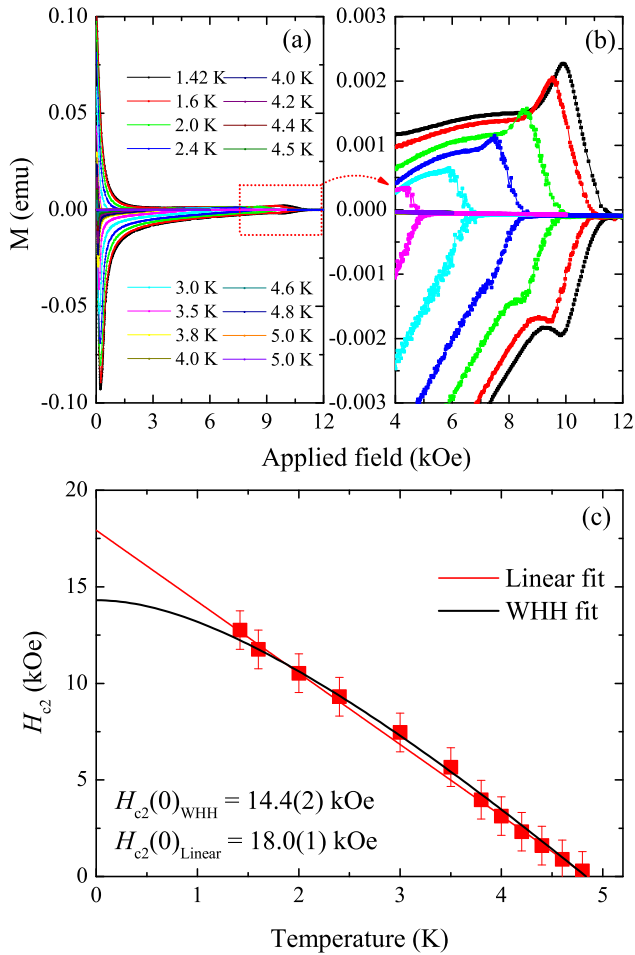}
\caption{(Color online) (a) $M$ vs. $H$ loops, collected at two different temperatures. (b) Magnified view of the data within the red dotted box in Fig.~\ref{fig:SrIrSn_Hc2}(a). (c) Temperature dependence of $H_{\rm c2}$ of Sr$_3$Ir$_4$Sn$_{13}$.}
\label{fig:SrIrSn_Hc2}
\end{figure}

Figure~\ref{fig:SrIrSn_Hc2}(a) shows  the first two quadrants of the $M$ vs. $H$ loops collected at different temperatures. Fig.~\ref{fig:SrIrSn_Hc2}(b) shows a magnification of the red dotted area in Fig.~\ref{fig:SrIrSn_Hc2}(a). A secondary peak or fish tail effect is detected in the magnetization loop at an applied field close to $H_{\rm c2}$. The peak effect slowly disappears as we move to the higher temperatures. Similar peak effects have also been observed in many other weak pinning superconductors~(see Ref.~[\onlinecite{Johansen}] and references therein). The presence of such a peak effect in Sr$_3$Ir$_4$Sn$_{13}$ may be a signature of additional pinning due to disorder. The temperature dependence of the upper critical field $H_{\rm c2}$ of Sr$_3$Ir$_4$Sn$_{13}$, determined from the point in the $M(H)$ loops where $M=0$, is shown in Fig.~\ref{fig:SrIrSn_Hc2}(c). $H_{\rm c2}(T)$ can be described using the Werthamer-Helfand-Hohenberg (WHH) model.~\cite{Werthamer,Helfand} The WHH fit yields $H_{\rm c2}=14.4(2)$ kOe at $T = 0$~K. A simple linear extrapolation of the data to $T=0$ K gives $H_{\rm c2}=18.0(1)$ kOe. \cite{  }$^{14}$

\begin{figure}[htb]
\includegraphics[width=1.0\linewidth]{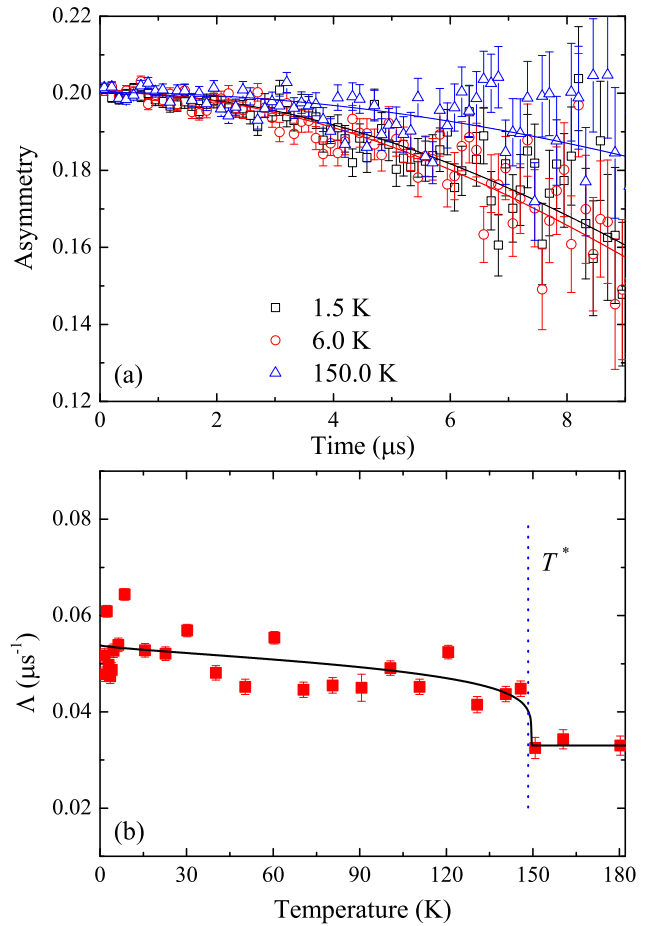}
\caption{(Color online) (a) ZF-$\mu$SR spectra of Sr$_3$Ir$_4$Sn$_{13}$ taken above (150 and 6 K) and below (at 1.5 K) $T_{\rm c}$. The solid lines are fits to the data using Eq.~\ref{eq:KT_ZFequation}. (b) Temperature dependence of the muon spin relaxation rate due to the presence of static nuclear moments. The solid line is a guide to the eyes.}
 \label{fig:SrIrSn_AsyZF}
\end{figure}

Figure~\ref{fig:SrIrSn_AsyZF}(a) compares the ZF-$\mu$SR signals collected above and below $T_{\rm c}$ and above $T^{*}$ at 150~K. The signals at 1.5 and 6.0~K are practically identical, implying that no additional magnetic moments (either static or dynamic) appear below $T_{\rm c}$. In contrast, the ZF-$\mu$SR signal taken at 150~K shows a lower relaxation rate. ZF-$\mu$SR data can be well described using a Gaussian Kubo-Toyabe relaxation function,~\cite{Kubo}
\begin{equation}
A(t)= A(0)\left\{\frac{1}{3}+\frac{2}{3}\left(1-\Lambda^2t^2\right){\rm exp}\left(-\frac{\Lambda^2t^2}{2}\right)\right\},
 \label{eq:KT_ZFequation}
\end{equation}
where $A(0)$ is the initial asymmetry and $\Lambda$ describes the muon spin relaxation rate due to the presence of static nuclear moments in Sr$_3$Ir$_4$Sn$_{13}$. The temperature dependence of $\Lambda$ is shown in Fig.~\ref{fig:SrIrSn_AsyZF}(b). The data show no apparent magnetic anomalies down to 1.5 K. However, a small but abrupt change in $\Lambda$ is observed at $\sim147$~K, indicating  a first-order type phase transition. Such a behavior is consistent with a structural transition or the onset of a CDW transition accompanied by strain or a lattice distortion. The muon probe is sensitive enough to detect such an effect, which arises due to the changes in the position of the muon stopping sites relative to the nuclear moments and hence the probed field distribution.

\begin{figure}[htb]
\includegraphics[width=1.0\linewidth]{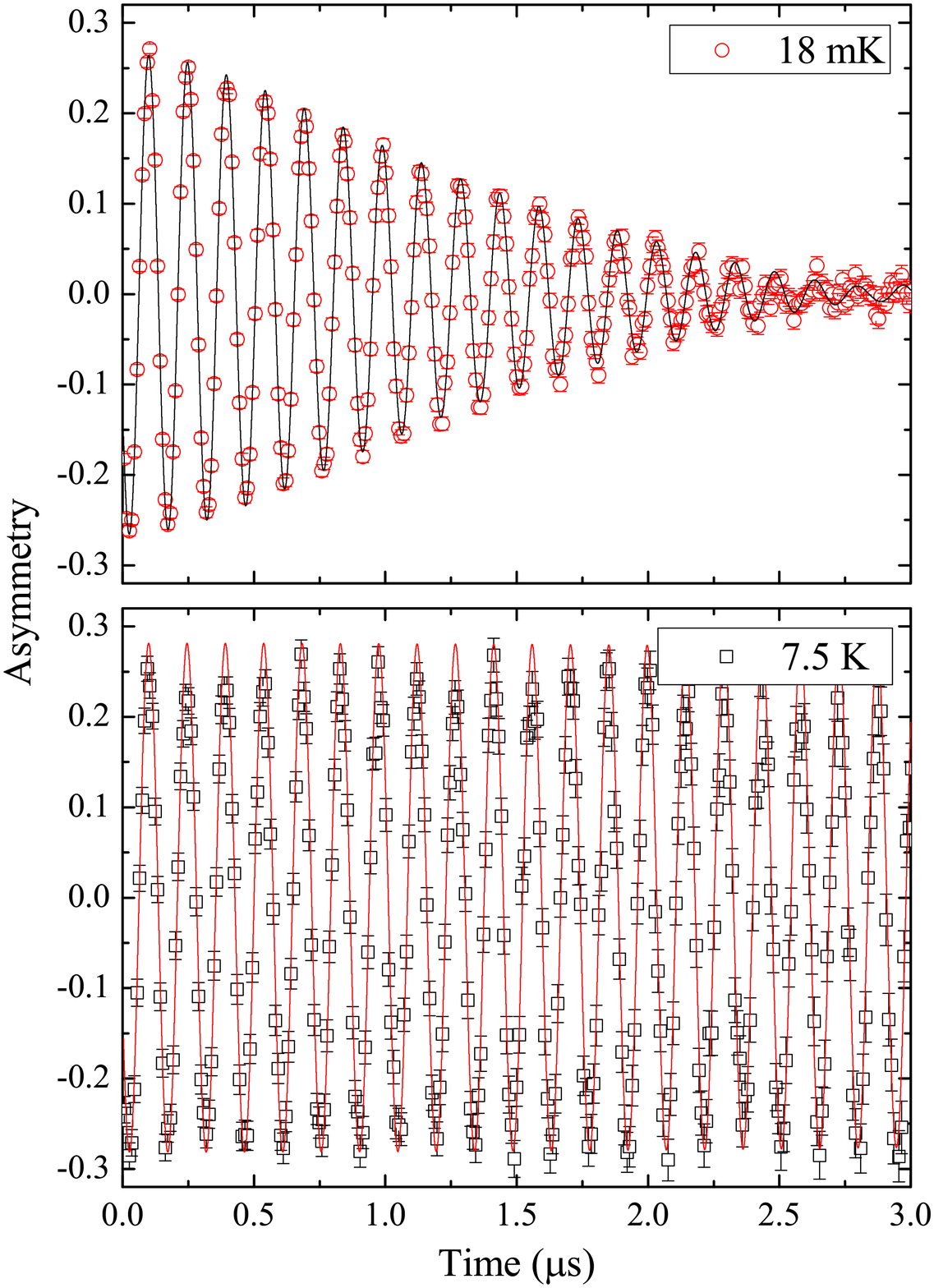}
\caption{(Color online) TF-$\mu$SR precession signals for Sr$_3$Ir$_4$Sn$_{13}$ measured above (at 7.5 K) and below (at 18 mK) $T_{\rm c}$ in an applied field of 500~Oe. The solid lines are the fits to the data using Eq.~\ref{eq:SM_Method}.}
 \label{fig:SrIrSn_AsyTF}
\end{figure}

Figure~\ref{fig:SrIrSn_AsyTF} shows the TF-$\mu$SR precession signals of Sr$_3$Ir$_4$Sn$_{13}$ taken both above (at 7.5 K) and below (at 18 mK) $T_{\rm c}$ in an applied field of 500~Oe. The signal in the normal state shows almost no damping, reflecting the homogeneous magnetic field distribution in the bulk of the material, whereas it decays very quickly below $T_{\rm c}$ due to the inhomogeneous field distribution generated by the superconducting vortex lattice.~\cite{Brandt}

\begin{figure}[htb]
\includegraphics[width=1.0\linewidth]{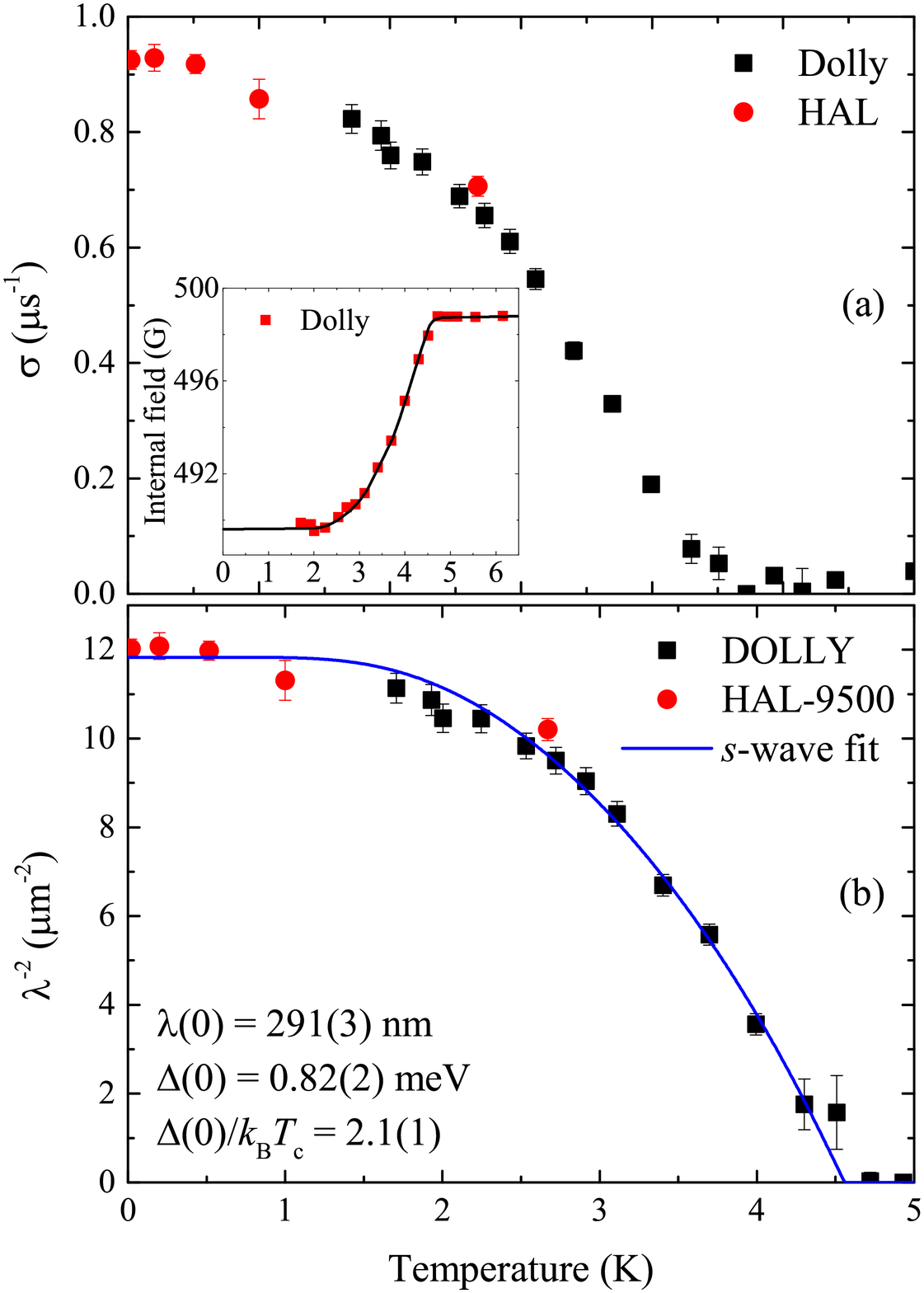}
\caption{(Color online) (a) Temperature dependence of the muon depolarization rate, $\sigma$ of Sr$_3$Ir$_4$Sn$_{13}$ collected on two spectrometers DOLLY and HAL-9500 at an applied magnetic field of 500 Oe. The inset shows the typical diamagnetic shift of the internal field experienced by the muons below $T_{\rm c}$. (b) The temperature dependence of $\lambda^{-2}(T)$. The solid line is a fit to the data using a single \textit{s}-wave BCS model.}
 \label{fig:SrIrSn_lambda}
\end{figure}

From the TF-$\mu$SR spectra we can determine the second moment of the magnetic field distribution associated with the vortex state and from this the magnetic penetration depth. Since the field distribution can be represented well by a multi-component Gaussian curve, the muon time spectra are fitted to a sum of $N$ Gaussian components:~\cite{Weber,Maisuradze}
\begin{multline}
A(t)=\sum^{N}_{i=1} A_{i}{\rm exp}(-\sigma^{2}_{i}t^{2}/2){\rm cos}(\gamma_{\mu}B_{i}t+\phi) \\
+A_{bg}{\rm cos}(\gamma_{\mu}B_{bg}t+\phi),
 \label{eq:SM_Method}
\end{multline}
where $\phi$, $A_{i}$, $\sigma_{i}$, and $B_{i}$ are the initial phase, asymmetry, relaxation rate, and mean field (first moment) of the $i$th Gaussian component, respectively. $A_{bg}$ and $B_{bg}$ are the asymmetry and field, respectively due to background contribution. We found that two Gaussian components ($N=2$) are sufficient to fit the muon time spectra data. For $N=2$, the first and second moments of $P(B)$ are given by
\begin{equation}
\left\langle B\right\rangle=\sum^{2}_{i=1} \frac{A_{i}B_{i}}{A_{1}+A_{2}},
 \label{eq:SM_1st_moment}
\end{equation}
and
\begin{equation}
\left\langle \Delta B^2\right\rangle=\frac{\sigma^{2}}{\gamma_{\mu}}=\sum^{2}_{i=1} \frac{A_{i}}{A_{1}+A_{2}}\left\{(\sigma_{i}/\gamma_{\mu})^2+[B_{i}-\left\langle B\right\rangle]^2\right\},
 \label{eq:SM_2nd_moment}
\end{equation}
where $\gamma_{\mu}=2\pi\times135.5388$ MHz/T is the muon gyromagnetic ratio and $\sigma$ the muon depolarization rate. Fig.~\ref{fig:SrIrSn_lambda}(a) shows the temperature dependence of $\sigma$ of Sr$_3$Ir$_4$Sn$_{13}$ for an applied field of 500 Oe. The inset shows the temperature dependence of the internal magnetic field at the muon site with the expected diamagnetic shift below $T_{\rm c}$. The solid line is a guide to the eye.

The superconducting contribution to $\sigma$ is obtained by subtracting the nuclear moment contribution (measured above $T_{\rm c}$) as ${\sigma_{sc}}^2=\sigma^2-{\sigma_{\rm nm}}^2$. In an isotropic type-II superconductor with a hexagonal Abrikosov vortex lattice described by Ginzburg-Landau theory, the magnetic penetration depth $\lambda$ is related to $\sigma
_{sc}$ by the equation:~\cite{Brandt}
\begin{equation}
\sigma_{sc}(b)[\mu{\rm s}^{-1}]=4.854\times10^4 (1 - b) [1 + 1.21(1 -\sqrt{b})^3]\lambda^{-2}[{\rm nm}^{-2}],
 \label{eq:Brandt_equation}
\end{equation}

Here $b=\left\langle B\right\rangle/B_{\rm c2}$ is a reduced magnetic field. In calculating $\lambda$, we have used the temperature dependence of $B_{\rm c2}=\mu_{0}H_{\rm c2}$ as shown in Fig.~\ref{fig:SrIrSn_Hc2}(c). Fig.~\ref{fig:SrIrSn_lambda}(b) shows the temperature dependence of $\lambda^{-2}$. Below 1~K, $\lambda^{-2}(T)$, which is proportional to the effective superfluid density, appears to flatten as is the case for fully gapped superconductors. This indicates that there are no nodes in the energy gap of Sr$_3$Ir$_4$Sn$_{13}$. The solid line in Fig.~\ref{fig:SrIrSn_lambda}(b) represents a fit to the data using the BCS  model:~\cite{Tinkham,Prozorov}
\begin{equation}
\frac{\lambda^{-2}(T)}{\lambda^{-2}(0)}= 1+2\int_{\Delta(T)}^{\infty}\left(\frac{\partial f}{\partial E}\right)\frac{E dE}{\sqrt{E^2-\Delta(T)^2}}.
 \label{eq:lambda}
\end{equation}
Here $\lambda^{-2}(0)$ is the  zero-temperature value of the magnetic penetration depth, and $f=[1+\exp(E/k_BT)]^{-1}$ is the Fermi function. Approximating the temperature dependence of the gap by~\cite{Carrington} $\Delta(T)=\Delta(0)\tanh\{1.82[1.018(T_c/T-1)]^{0.51}\}$, the fit yields $T_c=4.56(7)$~K, $\lambda(0)=291(3)$~nm, and $\Delta(0)=0.82(2)$~meV. The gap to $T_{\rm c}$ ratio $\Delta(0)/k_{\rm B}T_{\rm c}=2.1(1)$ is higher than the BCS value of 1.76, suggesting that Sr$_3$Ir$_4$Sn$_{13}$ is a strong-coupling superconductor. Using $H_{\rm c2}$ as determined from resistivity measurements $^{14}$ in Eq. \ref{eq:Brandt_equation} does not change these values
appreciably $\lambda(0)=311(3)$~nm, $\Delta(0)=0.79(2)$~meV and $\Delta(0)/k_{\rm B}T_{\rm c}=2.0(1)$. This is due to the fact that the applied field we used was only 500 Oe and that the correction in Eq. \ref{eq:Brandt_equation} is small for the relevant temperature range.

Using $H_{\rm c2}(0)=14.4(2)$~kOe and its relation with the coherence length $\xi$, $\left(\mu_{0}H_{\rm c2}=\frac{\phi_0}{2\pi\xi^2}\right)$, we calculate $\xi=15.1(2)$~nm at 0~K. This gives a $\kappa=\frac{\lambda}{\xi} \cong$ 19.

 By combining the value of $\xi$ and our measured value of $\lambda$, we can calculate the value of $H_{\rm c1}$ using the expression:~\cite{Brandt}
\begin{equation}
\mu_{0}H_{\rm c1}=\frac{\phi_0}{4\rm \pi\lambda^{2}}\left(\ln \frac{\lambda}{\xi}+0.5\right).
 \label{eq:hc1}
\end{equation}
We estimate $H_{\rm c1}(0)=60(1)$~Oe, which is in good agreement with the $H_{\rm c1}$ value of 81(1)~Oe, extracted from the magnetization measurements.
Our values of $\Delta(0)$ and gap-to-$T_{\rm c}$ ratio obtained by a microscopic measurements are in good agreement with those obtained from specific heat measurements,~\cite{Kase} whereas $\xi(0)$ is slightly larger (and $\lambda(0)$ smaller) than in Ref.~\onlinecite{Kase}, indicating a cleaner material.


In conclusion, magnetization and $\mu$SR measurements have been performed on superconducting Sr$_3$Ir$_4$Sn$_{13}$. From the magnetization measurements we determine the temperature dependences of the lower and upper critical fields. ZF-$\mu$SR results do not find evidence of any magnetism in Sr$_3$Ir$_4$Sn$_{13}$, but shows at $T^*$ a small increase in the muon spin relaxation rate consistent with a structural and/or CDW scenario accompanied by a lattice distortion.

TF-$\mu$SR results show that the superfluid density, $\rho_s\propto\lambda^{-2}$ becomes temperature independent at low temperature, which is consistent with a fully gapped superconducting state, with $\rho_s(T)$ well described within the single $s$-wave gap scenario with $\Delta(0)=0.82(2)$~meV and penetration depth, $\lambda(0)=291(3)$~nm. There is no signature of multiple gaps, as it may be expected from the multiband structure of this material possibly indicating that the gaps have a similar magnitude. The value of the gap to $T_{\rm c}$ ratio, 2.1(1) is higher than the BCS value of 1.76 and suggests that Sr$_3$Ir$_4$Sn$_{13}$ is a strong-coupling superconductor.  The results presented here will provide a reference point for studies under pressure to detect any changes of the superconducting and magnetic properties in this and related material. The prospect of suppressing the CDW-gap and increasing $T_{\rm c}$ under hydrostatic pressure motivates further $\mu$SR as well as other studies under pressure to understand the interplay between $T_{\rm c}$ and $T^*$ in the ternary intermetallic stannides.

The $\mu$SR experiments were performed at the Swiss  Muon Source (S$\mu$S), Paul Scherrer Institute (PSI, Switzerland). Work at Brookhaven is supported by the US DOE under Contract No. DE-AC02-98CH10886.


\begin{thebibliography}{99}
\bibitem{  } Resistivity versus magnetic field measurements at fixed temperatures on this sample give a higher $H_{\rm c2}(0)$=37 kOe. However, such a determination may be influenced by impurity phases.
\end{thebibliography}
\end{document}